\documentclass[12pt,tightenlines]{revtex4} 
\usepackage{color}
\usepackage{graphicx}
\usepackage[small,bf]{caption}
\usepackage{epsfig,wrapfig,sidecap}
\usepackage{fancyhdr,lastpage}
\usepackage[T1]{fontenc}
\usepackage[latin1]{inputenc}
\usepackage[english]{babel}
\usepackage{helvet}

\newcommand{\shorttitle}{Computing Frontier: Cosmic Frontier Science}
\newcommand{\projectnum}{}
\newcommand{\piname}{}

\newcommand{\smarial}[1]{\fontsize{10pt}{0pt} \em{#1}}

\begin{document}

\begin{center}
{\bf\scshape{Snowmass Computing Frontier: Computing for the Cosmic Frontier,
    Astrophysics, and Cosmology}}
\end{center}

\begin{center}
{Andrew Connolly$^1$, Salman Habib$^2$, Alex Szalay$^3$; Julian Borrill$^4$, George
Fuller$^5$,}\\
{Nick Gnedin$^6$, Katrin Heitmann$^2$, Danny Jacobs$^7$, Don Lamb$^8$, Tony
Mezzacappa$^9$,}\\
{Bronson Messer$^9$, Steve Myers$^{10}$, Brian Nord$^6$, Peter Nugent$^4$,
Brian O'Shea$^{11}$,}\\ 
{Paul Ricker$^{12}$, Michael Schneider$^{13}$}
\end{center}

\begin{center}
$^1$University of Washington, Seattle, WA\\
$^2$Argonne National Laboratory, Lemont, IL\\
$^3$Johns Hopkins University, Baltimore, MD\\
$^4$Lawrence Berkeley National Laboratory, Berkeley, CA\\
$^5$UC San Diego, La Jolla, CA\\
$^6$Fermi National Accelerator Laboratory, Batavia, IL\\
$^7$Arizona State University, Tempe, AZ\\
$^8$University of Chicago, Chicago, IL\\
$^9$Oak Ridge National Laboratory, Oak Ridge, TN\\
$^{10}$National Radio Astronomy Observatory, Socorro, NM\\
$^{11}$Michigan State University, East Lansing, MI\\
$^{12}$University of Illinois, Urbana-Champaign, IL\\
$^{13}$Lawrence Livermore National Laboratory, Livermore, CA
\end{center}

\medskip

{\center{\bf Abstract:} {This document presents (off-line) computing
    requrements and challenges for Cosmic Frontier science, covering
    the areas of data management, analysis, and simulations. We invite
    contributions to extend the range of covered topics and to enhance
    the current descriptions.}}

\section{Introduction}
\label{sec:comp-intro}

The unique importance of the ``Cosmic Frontier'', the interface
between particle physics, cosmology, and astrophysics has been long
recognized. With the cementing of the cosmological ``Standard Model''
-- well measured, but deeply mysterious -- recent progress in this
area has been dramatic, and has propelled the field to the forefront
of research in fundamental physics. Topics of key interest in the
Cosmic Frontier include dark energy, dark matter, astrophysical and
cosmological probes of fundamental physics, and the physics of the
early universe. Research in these topics impacts our basic notions of
space and time, the uneasy interaction between gravity and quantum
mechanics, the origin of primordial fluctuations responsible for all
structure in the Universe, and provides unique discovery channels for
reaching beyond the Standard Model of particle physics in the dark
matter and neutrino sectors. Excitingly, a host of new experiments and
observations are poised to significantly advance, perhaps decisively,
our current understanding of the Cosmic Frontier.

Experimental and observational activities in the Cosmic Frontier cover
laboratory experiments as well as multi-band observations of the
quiescent and transient sky. Direct dark matter search experiments,
laboratory tests of gravity theories, and accelerator dark matter
searches fall into the first class. Investigations of dark energy,
indirect dark matter detection, and studies of primordial fluctuations
fall into the second class; essentially the entire set of available
frequency bands is exploited, from the radio to TeV
energies. Relevant theoretical research also casts a very wide net --
ranging from quantum gravity to the astrophysics of galaxy formation.

The size and complexity of Cosmic Frontier experiments is diverse (see
Table~I), spanning tabletop experiments to large
scale surveys; simultaneously covering precision measurements and
discovery-oriented searches. A defining characteristic of the Cosmic
Frontier is a trend towards ever larger and more complex experimental
and observational campaigns, with sky surveys now reaching
collaboration memberships of over a thousand researchers (roughly the
same size as a large high energy physics
experiment). Cross-correlation of these experimental datasets enables
the extraction of more information, helps to eliminate degeneracies,
and reduces systematic errors. These factors count among the major
drivers for the computational and data requirements that we consider
below.
\begin{table}[h]
\begin{center}
\begin{tabular}{|l|c|c|c|} 
 \hline 
{\bf Experimental Data} & 2013 & 2020 & 2030+ \\
\hline
Storage & 1PB & 6PB & 100-1500PB \\
Cores & 10$^3$ & 70K & 300+K \\
CPU hours & 3x10$^6$ hrs & $2\times 10^8$ hrs & $\sim 10^9$ hrs \\
 \hline 
{\bf Simulations}  & 2013 & 2020 & 2030+ \\
 \hline 
 Storage & 1-10 PB & 10-100PB & $> 100$PB - 1EB\\
Cores & 0.1-1M & 10-100M &$> 1$G\\
CPU hours & 200M & $>$20G & $> 100$G\\
\hline
\end{tabular}
\caption{Estimated compute and storage needs for the next 10-20 years
  of Cosmic Frontiers simulations and experiments..}
\end{center}
\end{table}

\subsection{Computing at the Cosmic Frontier}

The role of computing in the Cosmic Frontier is two-fold: on-line
processing at the instrument, including its data management system,
and off-line processing for science. The off-line processing includes
all related simulation activity as well as data archiving and
curation. Because the nature of the detectors varies widely we do not
consider instrument-specific (`front-end') processing as being within
the purview of this document. We will focus on general features of the
data management and analysis chain and on the associated off-line
computing requirements.

\section{Cosmic Frontier: Facilities}
\subsection{Experimental Facilities}

The dramatic increase in data from Cosmic Frontier experiments over
the last decade has led to a number of fundamental breakthroughs in
our knowledge of the ``Dark Universe'' and physics at very high
energies. Driven by advances in detectors and instruments, current
experiments generate on the order of a petabyte of data per
year. Projecting the growth in data over the next two decades, based
on the expected technological capabilities of the next generation of
experiments, it is expected that there will be an increase of between
a factor of 100 and 1000 in data volumes by 2035. This will see the
Cosmic Frontier experiments reach an exabyte of data. While the size
of data from individual experiments is driven by cosmological surveys,
the rate of increase is common to many of the Cosmic Frontier missions
-- from direct dark matter detection to the characterization of
neutrino masses. This growth in data and in its complexity presents a
number of computational challenges that must be addressed if we wish
to properly exploit the scientific and discovery potential of these
experiments.

\subsubsection{Optical Surveys}

Current Cosmic Frontier experiments survey thousands of square degrees
of the sky using large format CCD cameras and wide field multi-object
spectrographs. The largest current DOE/NSF imaging survey is the Dark
Energy Survey (DES); a multi-band optical survey of
$\sim$5000 deg$^{2}$ of the southern sky that utilizes the 4m Blanco
telescope at CTIO and a 570M pixel camera. With survey operations
starting in 2013, DES is expected to operate for 525 nights and to
generate approximately one petabyte of data. Designed as a Stage III
dark energy experiment it will address the nature of dark energy
through multiple cosmological probes including measurements of weak
gravitational lensing, the angular clustering of galaxies, number
densities of galaxy clusters, and luminosity distances for
supernovae. 

The spectroscopic counterpart to DES is BOSS (the Baryon Oscillation
Spectroscopic Survey)~\cite{boss}; a multi-fiber spectroscopic survey
that utilizes the Sloan Digital Sky Survey telescope (as part of
SDSS-III) to measure the redshifts for 1.3M luminous galaxies with
$z<0.7$ and the clustering of the Lyman-$\alpha$ forest for $>160,000$
high redshift quasars with $z>2$. The BOSS survey started operations
in 2009 and is expected to complete its observations in 2014 but will
likely be extended (as eBOSS~\cite{eboss}) as part of SDSS-IV and will
run from 2014 to 2017. Following eBOSS is the Dark Energy
Spectroscopic Instrument (DESI)~\cite{desi}, a Stage IV dark energy
experiment, comprising of a multi-fiber spectrograph on the Mayall 4-m
telescope. Slated to begin survey operations in 2018, DESI fits in
between the end and beginning of two imaging surveys, DES and the
Large Synoptic Survey telescope (LSST), and will obtain redshifts for
more than 20 million galaxies and quasars. Through a combination of
baryon acoustic oscillations (BAO) measurements and measurements of
redshift space distortions (RSD), eBOSS (Stage III), and DESI will
achieve accuracies of better than 1\% on measures of the angular
diameter distances (with DESI producing these measurements at $z<2$ with
3-D galaxy maps and at $z>2$ with the Lyman-$\alpha$ forest, in 35
redshift bins).

Over the next two decades the size of dark energy surveys, together
with the volumes they probe, will grow by two orders of magnitude.
Ground- and space- based Stage IV imaging surveys (e.g., HSC (Hyper
Suprime-Cam)~\cite{hsc}; LSST\cite{lsst}, Euclid\cite{euclid}, and
WFIRST (Wide Field Infra Red Survey Telescope)~\cite{wfirst}) will
survey the high redshift and temporal Universe at unprecedented
resolution.  The largest of the imaging surveys, LSST is a joint
DOE-NSF led initiative. As a Stage IV dark energy mission it will
survey $>$18,000 deg$^{2}$ of the sky $\sim 1000$ times over the
period of 10 years. Starting in 2020 it will amass $>60$PB of imaging
by the completion of operations in 2030.

Complementing the optical ground-based telescopes, the joint ESA led
space-based Euclid mission (with a 1.2m Korsch telescope) will survey
15,000 deg$^{2}$ of the sky in a single broad optical filter and 3
near-infrared photometric passbands ($1-2\mu$m). The imaging survey
will be supplemented with a 15,000 deg$^{2}$ near-infrared
spectroscopic survey that will measure 40 million redshifts with an
accuracy of $z < 0.001(1+z)$.  Euclid is expected to launch in 2020
and have a survey mission lifespan of approximately 6.25 years. Euclid
will likely be followed in 2023 by the NASA-led WFIRST space
mission. This Stage IV dark energy mission, while still under
definition, is expected to comprise a 2.4m aperture telescope and a
large IR camera (with 0.11-arcsec pixels) that will provide
spectroscopic and imaging over its 6 year mission lifetime.

Beyond the DOE sponsored experiments described previously, this growth
in imaging data will be matched by a 30-fold increase in spectroscopic
observations of distant galaxies (in addition to the near-infrared
spectroscopy available from Euclid and WFIRST). HETDEX (Hobby-Eberly
Telescope Dark Energy Experiment)~\cite{hetdex}, and the Prime Focus
Spectrograph (PFS \cite{pfs}), comprise fiber spectrographs with the
capability of intermediate resolution spectroscopy of a several
thousand simultaneous targets (with up to 33,500 fibers for
HETDEX). Surveying between 300 and 18,000 deg$^{2}$ over the period
2014--2023 these missions will result in over 30 million galaxies with
spectroscopic redshifts by the mid-point of the LSST mission.

While focused on probes of dark energy and dark matter the datasets
generated by these surveys will have a much broader reach in terms the
cosmological and astrophysical questions they will address (e.g., the
sum of the masses of neutrinos). The increase in data volume, the
complexity of the data, and the sensitivities of these surveys to
systematic uncertainties within the data are all driving the
computational complexity and needs of the Cosmic Frontier in
2020+. The computational requirements for processing of these data
alone will increase by 2-3 orders of magnitudes by the middle of the
next decade (driven primarily by cosmological imaging surveys). While
the individual dark energy missions will provision the computational
resources appropriate to process their data as it is collected,
project-led processing must meet the exacting requirements of dark
energy science (particularly as the computational requirements grow
quadratically with time due to the reprocessing required by sequential
data releases). Facilities that possess the capability of reanalyzing
petascale data resources throughout the lifespan of these missions
will be required. As an illustrative example, reprocessing the LSST
images five years into the survey will require approximately a billion
CPU-hours.

\subsubsection{Cosmic Microwave Background
      Surveys}

Cosmic Microwave Background (CMB) surveys have been at the forefront
of observational cosmology for over two decades. The next generation
of CMB experiments will focus on measurements of the polarization of
the CMB as a probe of inflationary physics. For example, ``B-mode''
polarization is a distinctive signature and probe of the gravitational
waves that were generated during the inflationary period in the early
universe and gravitational lensing of the CMB directly constrains the
sum of the masses of neutrino species. The goals of Stage IV CMB
experiments (CMB-S4) are to improve the resolution and
signal-to-noise of CMB polarization measurements by increasing the
number of detectors to $\sim$500,000 (a 30-fold increase over Stage
III CMB experiments). This will enable micro to nano-Kelvin
measurements of the CMB on scales of arc-minutes (thereby
constraining, for example, the masses of neutrinos to accuracies of
10-15 meV). Based on current projections, CMB-S4 (situated at multiple
sites to provide large area survey capabilities) is expected to
achieve first light at the beginning of the next decade.

The computational requirements associated with these next generation
CMB experiments follow the exponential growth in data over the last
decade.  Sampling at 100 Hz and for a 5 year survey CMB-S4 will result
in over 10$^{15}$ time samples of the CMB sky and maps of
$\sim$10$^{9}$ sky pixels (with 10,000 times the number of
observations per pixel than current CMB missions). At this scale,
map-making, foreground removal, and power-spectrum estimation
techniques (including estimates of the covariance associated with
these measures) become computationally challenging. Algorithms must
scale no more than linearly in the number of time samples. One
consequence of this is the need to use Monte Carlo methods (MC) for
debiasing and uncertainty quantification, resulting in analyses that
scale linearly in the numbers of MC realizations, iterations and
samples. Assuming O(10$^4$) realizations (for 1\% uncertainties) and
O(10$^2$) iterations using current methods, next-generation
experiments will require up to O(10$^{21}$) operations. Reducing this
number is an active area of research.

\subsubsection{Radio Surveys}

NRAO has recently announced the VLA Sky Survey (VLASS)~\cite{vlass}
initiative to plan and prosecute a new suite of extensive radio
synoptic sky surveys carried out with the recently completed upgrade
of the VLA~\cite{evla}. The success of a VLASS will be contingent on
real-time and prompt processing and imaging of the data, and will be a
key opportunity to put into practice new algorithms developed to deal
with the large data rates that are now possible. This survey will span
a decade and will pave the way and significantly complement surveys
such as SDSS, Pan-STARRs and LSST. The data management requirements of
the VLASS in the fast-wide imaging area are currently a high-risk
area, and common solutions with those needed for LSST would be of
great utility to this project.

A VLASS will produce an overall image of up to the 30000 square
degrees of the visible sky in multiple epochs in the 1000 or more
frequency channels. This image data cube, at an angular resolution of
0.1 arcseconds, contains nearly 40 petapixel per epoch and must be
stored, processed, studied, and disseminated. In particular, the
cross-correlation and comparison with views of the same sky from other
facilities and missions such as Planck, Spitzer, JWST (James Webb
Space Telescope)~\cite{jwst}, and LSST will need to be routine and
accessible in order to carry out critical science goals such as the
astrophysics and cosmology from large-scale structure mapping.

The Square Kilometre Array (SKA)~\cite{ska} project is an
international program to construct the next-generation of large radio
arrays operating at frequencies from 50MHz to 10GHz. Although the US
has currently not provided funding to the SKA in its current phase,
paticipation will be critical in order to have access to the SKA data
and products for key science aspirations, such as the use of radio
weak lensing measurements for dark energy studies as outlined in the
DETF (Dark Energy Task Force) report~\cite{detf}. Even in the SKA
Phase 1 that is currently under development this facility will require
support for data rates greatly exceeding those for the VLA
and HERA (HEterodyne Receiver Array)~\cite{hera} projects. Work that
would be carried out for pathfinders such as the VLA will be directly
applicable to SKA planning, and can be considered to form a
substantial part of the portfolio that the US can bring to the table
later in the decade. SKA precursor instruments are already taking
data, for example the Murchison Widefield Array (MWA)~\cite{mwa} is
accumulating data at 1PB/year.

\subsubsection{Direct Dark Matter Detection
      Experiments}

Direct detection experiments address the nature of dark matter through
the identification of interactions of Weakly Interacting Massive
Particles (WIMPs) with matter.  The sensitivity of these experiments
has increased dramatically over the last decade due to increases in
the size and complexity of the detectors and improvements in the
suppression of background signals~\cite{dmrev}. While smaller in scale
than dark energy experiments, second generation (G2) experiments are
expected to see a comparable order of magnitude increase in the size
of detectors and in the volume of data they will generate. By the end
of this decade direct dark matter experiments will exceed petabyte
sized datasets.  This increase in scale of data will require a
comparable increase in the sophistication and complexity of software
frameworks used in processing these data (including the development of
the infrastructure to manage the data, metadata, and analysis codes).

Throughout the development of dark matter detection experiments,
simulations have played an integral role in defining the properties
and sensitivities of the instruments and backgrounds (e.g., Fluka,
Geant4, MCNPX, MUSUN, MUSIC and SOURCES). Over the coming decade, with
the increase in sensitivity of G2 and beyond, the importance of
simulations will grow and the development of the simulation and
analysis frameworks will need to be coordinated across experiments in
order to maximize the efficient use of resources. HEP is a natural
organization to foster and manage such collaborations.

\subsubsection{Impact of Technology Developments}

Current experiments, such as DES, will generate a petabyte of data
over the period 2013-2017. Next-generation surveys such as CMB-S4 or
LSST will increase these data volumes to $15-100$PB on the 2020-2030
timescale. Experiments such as LIGO (Laser Interferometer
Gravitational-Wave Observatory)~\cite{ligo} with a ``multi-messenger''
science program will increase the complexity of the data processing
environments.  Computational requirements for processing, and
archiving these data are projected to grow from $\sim$70K cores in
2020 to $\sim$280K cores by 2030 (here a core is defined as a core in
a modern multi-core CPU). Post-2030 technology trends, including
energy resolving detectors such as Microwave Kinetic Inductance
Detectors (MKIDs) and advances in radio detectors, are expected to
maintain the steep growth in data. The near-future VLASS will be
followed by the SKA expected to come on-line in the next decade and to
generate between 300 and 1500PB of data per year.

\subsection{Simulation Facilities}

The intrinsically observational nature of much of Cosmic Frontier
science implies a great reliance on simulation and modeling. Not only
must simulations provide robust predictions for observations, they are
also essential in planning and optimizing surveys, and in estimating
errors, especially in the nonlinear domains of structure
formation. Synthetic sky catalogs play important roles in testing and
optimizing data management and analysis pipelines. The scale of the
required simulations varies from medium-scale campaigns for
determining covariance matrices to state of the art simulations for
simulating large-volume surveys, or, at the opposite extreme,
investigating dark matter annihilation signals from dwarf galaxies.

Required facilities for carrying out simulations include large-scale
supercomputing resources at DOE and NSF National Centers, local
clusters, and data-intensive computing resources needed to deal with
the enormous data streams generated by cosmological simulations. The
data throughput can, in many cases, easily exceed that of
observations; data storage, archiving, and analysis requirements
(often in concert with observational data) are just as demanding as
for observational datasets. Simulation data analysis is expected to be
carried out in a two-stage manner, the first stage close to where the
data is generated, and the second at remote analysis sites. Data
transfer rates associated with this mode of operation are likely to
stress the available network bandwidths; future requirements in this
area have not yet been fully spelled out.

In terms of computing resources, although there are significant
challenges in fully exploiting future supercomputing hardware,
resources expected to be available should satisfy simulation
requirements, currently at the 10PFlops scale and expected to reach
the exascale after 2020. The data-related issues are more serious and
will need changes in the current large-scale computing model, as
covered in more detail in Section~4. Successful implementation of the
recently suggested Virtual Data Facility (VDF) capability at ALCF
(Argonne), NERSC (LBNL), and OLCF (Oak Ridge), would go a long way
towards addressing these issues for Cosmic Frontier simulations as
well as for analysis of observational datasets.

\section{Simulations}

Simulations cover a broad area ranging from predictions for various
theoretical scenarios and for characterizing cosmological probes, to
detailed end-to-end simulations for specific experiments, to the Monte
Carlo-based analyses for extracting cosmological and other parameters
from observational data.

\subsection{Role of Simulations}

It is widely recognized that simulation plays a critical role in
Cosmic Frontier science, not only as the primary tool for theoretical
predictions, but even more significantly, in evaluating and
interpreting the capabilities of current and planned experiments. For
optical surveys, the chain begins with a large cosmological simulation
into which galaxies and quasars (along with their individual
properties) are placed using semi-analytic or halo-based models. A
synthetic sky is then created by adding realistic object images and
colors and by including the local solar and galactic
environment. Propagation of this sky through the atmosphere, the
telescope optics, detector electronics, and the data management and
analysis systems constitutes an end-to-end simulation of the survey. A
sufficiently detailed simulation of this type can serve a large number
of purposes such as identifying possible sources of systematic errors
and investigating strategies for correcting them and for optimizing
survey design (in area, depth, and cadence). The effects of systematic
errors on the analysis of the data can also be investigated; given the
very low level of statistical errors in current and next-generation
precision cosmology experiments, and the precision with which
deviations from $\Lambda$CDM are to be measured, this is an absolutely
essential task.

Simulations at smaller scales are important for indirect dark matter
detection studies and for understanding astrophysical systematics in
structure formation-based probes of cosmic structure. The latter
includes baryonic effects and the role of intrinsic alignments in weak
lensing shear measurements. At even smaller scales, simulations are
very important in identifying and understanding possible systematics
in the use of Type 1a supernovae as standard candles, and in
identifying neutrino property signatures in core collapse supernovae.

Directly analogous to the situation in building a community-supported
software base for Cosmic Frontier experiments, there is a related need
for bringing together larger collaborations in the area of
simulations. The Lattice QCD community has shown what is possible in
this direction by working together in a cohesive manner. Although
consolidation within Cosmic Frontier computing is much more difficult
due to the large number of individual efforts in different fields,
initial attempts are showing some promise and will hopefully come to
fruition in the near term.

\subsection{Computational Challenges}

We now enumerate a number of computational challenges faced by the
Cosmic Frontier simulation community in achieving the stated science
goals of next generation observations. These simulations span a wide
variety of techniques, from N-body simulations, to gasdynamics, and
finally to fully consistent quantum transport.

\subsubsection{N-body Simulations}

Cosmological simulations can be classified into two types: (i)
gravity-only N-body simulations, and (ii) hydrodynamic simulations
that also incorporate gasdynamics, sub-grid modeling, and feedback
effects~\cite{dolag}. Because gravity dominates on large scales, and
dark matter outweighs baryons by roughly a factor of five, N-body
simulations provide the bedrock on which all other techniques
rest. These simulations accurately describe matter clustering well out
into the nonlinear regime, possess a wide dynamic range (Gpc to kpc,
allowing coverage of survey-size volumes), have no free parameters,
and can reach sub-percent accuracies. Much of our current knowledge of
nonlinear structure formation has been a direct byproduct of advances
in N-body techniques. In the foreseeable future, N-body simulations
will continue to remain the workhorses (and racehorses!) of
computational cosmology.

The key shortcoming of this approach is that the physics of the
baryonic sector cannot be treated directly. To address this problem,
insofar as galaxy surveys are concerned, several post-processing
strategies exist. These incorporate additional modeling and/or physics
on top of the basic N-body simulation, e.g., halo occupation
distribution (HOD), sub-halo abundance matching (SHAM), and
semi-analytic modeling (SAM) for incorporating galaxies in the
simulations. Nevertheless, there is a significant current gap in our
ability to incorporate baryonic physics, as gleaned from hydro
simulations, into N-body codes~\cite{motext}. Were this possibility to
be more fully realized than in the current state of the art, the
ability to control systematic errors would be significantly enhanced.

As discussed already, large-scale cosmological N-body codes are
essential for the success of all future cosmological surveys. The
fundamental problem in addressing the needs of the surveys is not only
to run very large simulations at demanding values of force and mass
resolution, but also a large number of them. At the same time,
analysis and post-processing of the simulation data is in itself a
challenging task (Section 4), since this data stream can easily exceed
that of observations.

Parallel numerical implementations of N-body algorithms can be purely
particle-based or employ particle/grid hybrids. For survey
simulations, a spatial dynamic range of about a million to one is
needed over the entire simulation box, while the dynamic range in mass
resolution, in terms of the mass ratio of the most massive object to
the lowest mass halo resolved in the simulation is similar, perhaps an
order of magnitude less. For simulations that study individual
galaxies and clusters (e.g., for indirect dark matter detection
studies), the spatial dynamic range remains the same, but the
computational problem becomes much more severe, because natural
time-scales become significantly smaller. As supercomputer
architectures evolve in more challenging directions, it is essential
to develop a powerful next generation of these codes that can
simultaneously avail various types of hybrid and heterogeneous
architectures.

\subsubsection{Hydrodynamic Simulations}

The increased focus on precision cosmological measurements over the
past decade has led to a regime in which controlling systematic
errors is an overarching concern. Because baryonic processes are
typically among the leading systematic uncertainties in measurements
of halo structure, abundance, and clustering properties, this concern
has driven simulators to push the development of improved methods for
handling baryons. This development broadly falls into two categories:
new techniques for directly solving the hydrodynamic equations
together with additional terms to handle magnetic fields, nonthermal
relativistic species, conduction, and viscosity; and subresolution
models for stellar and black hole feedback. Accompanying this
development has been an increasing need for observational validation
of these techniques and models via numerical studies of
observationally well-resolved nearby objects in the Local Group and
beyond. These trends reflect the larger evolution of cosmology from a
data-starved to a data-rich science.

The most crucial fact about hydrodynamic cosmological simulations is
that neither in this decade nor, most likely, in the next, will they
reach the precision of N-body simulations -- not because of the
numerical difficulties, but because the physics that must be modeled
is too complex and/or still poorly understood. How to correctly couple
unresolved, sub-grid physics like star formation, supernovae feedback,
and black hole feedback (including SMBH feedback) to cosmological
simulations remains an unsolved problem.  This is a formidable
challenge for hydrodynamical simulations, and solving it will be
crucial to addressing a number of critical topics in galaxy formation.
Hence, it appears that in the foreseeable future, ``baryonic
effects'', i.e.\ effects of baryonic physics (including hydrodynamics
and other gas and stellar physics) on the probes of dark matter and
dark energy will remain an effective ``systematic error'' that must be
calibrated out or marginalized over.

The primary role of hydrodynamic simulations in cosmology is then to
provide more or less realistic examples of baryonic effects on various
probes of large-scale structure (weak and strong lensing, galaxy
clustering, matter-galaxies cross-correlations, redshift-space
distortions, etc). These examples will serve as the basis for
phenomenological models that parameterize baryonic effects in some
form (for example, nuisance parameters) and that can be used directly
in constraining the dark sector.

Two exceptions to that general approach are possible. The first one is
modeling of cosmic reionization~\cite{cosreion}. The process of
reionization leads to the transition from mostly neutral to a highly
ionized state for most of the cosmic gas. This ionized gas serves as a
semi-transparent screen for the CMB, affecting CMB fluctuations in a
well understood manner. Since no future competitive cosmological
constraints are possible without using the CMB, reaching the maximally
achievable precision from the CMB data is crucially
important. However, since the effect of reionization on the CMB is not
large (less than 10\%), simulations of reionization are not required
to be particularly accurate (a 10\% simulation error results in less
than 1\% error in extracted cosmological parameters). In addition, the
currently deployed petascale supercomputing platforms are the ideal
target for simulations of cosmic reionization. Hence, it is likely
that by the end of this decade, direct numerical simulations of cosmic
reionization will reach the precision that, in relative terms (i.e. in
terms of the precision on extracted cosmological parameters) will be
commensurate with the accuracy of N-body simulations.

A second exception is modeling of clusters of
galaxies~\cite{clusters}. Clusters remain an important cosmological
probe, both of dark energy and of modified gravity, and a substantial
effort is currently expended on improving numerical and physical
modeling of clusters. Just as in the previous case, the actual
precision of hydrodynamic simulations will remain well below that of
N-body simulations, primarily due to uncertainties in modeling the
physics of AGN feedback in clusters.  However, clusters are extremely
rare objects, sitting at the very tail of the exponential
distribution, and therefore the sensitivity of their properties and
abundance to cosmological parameters is extremely high. Hence, a
relatively low precision of future hydrodynamic simulations can be
compensated by the high sensitivity of clusters as cosmological
probes. The main challenge in the next decade will be to improve the
fidelity of cluster simulations, particularly the physics of stellar
and AGN feedback, to the level where the cosmological constraints from
cluster observations are competitive with other cosmological probes.

\subsubsection{ Emulators}

Future cosmological surveys will probe deep into the nonlinear regime
of structure formation. Precision predictions in this regime will be
essential to extract the cosmological information contained on these
scales, as well as to control systematic errors at larger length
scales via cross-validation techniques. Such predictions can only be
obtained from detailed simulations over a wide range of cosmological
models, augmented by observational inputs. These simulations are
computationally very expensive; it is imperative to develop a strategy
that allows precise predictions from only a limited number of such
simulations.

It has been shown that accurate prediction tools, so-called emulators,
can be built from a (relatively) limited set of high-quality
simulations~\cite{emu}. Building such emulators relies on optimal
strategies for generating simulation campaigns in order to explore a
range of different cosmological models. The emulators are then
constructed using sophisticated interpolation schemes (Gaussian
Process modeling is one example). They replace the expensive
simulators as predictors for observables within the parameter space to
be explored and are used for calibration against data via MCMC methods
to determine parameter constraints. While it has been demonstrated
that the general idea works extremely well and prediction tools at the
percent level accuracy can be generated, this line of work has to be
extended in several crucial ways with the advent of future
surveys. Examples of such extensions include but are not limited to:
(i) a broader set of measurements -- going beyond the current focus on
observables such as the matter power spectrum and the halo
concentration mass relation, to many more, e.g. the halo mass
function, galaxy correlation functions, etc.; (ii) a broader set of
cosmological models, extending beyond wCDM models; (iii) inclusion of
baryonic effects; (iv) determination of covariances (see below).

\subsubsection{Covariance Matrices}

Determining cosmological parameters requires not only the accurate
predictions of the observables, but also the \emph{errors} on those
observables. Cosmological observables are commonly multivariate
(e.g. correlation functions) requiring the specification of correlated
errors in the form of covariance matrices. The correlated errors
between multiple types of cosmological probes are equally
important. This error estimation is computationally demanding and has
important implications for parameter inference given finite computing
resources.

Errors in the sample covariance estimator due to finite $N$ propagate
($N$ is the number of samples) into increased uncertainties in the
cosmological parameters that scale with the ratio $\sqrt{N_b/N}$,
where $N_b$ is the number of `bins' in the multivariate observable. To
restrict the degradation of cosmological parameters to no more than,
e.g., 10\%, one quickly obtains requirements of $\sim10^{4}-10^{6}$
expensive cosmological simulations that need to be run for next
generation surveys~\cite{cov}. The requirements for
cosmological covariance estimation are even more demanding than the
above estimate when the errors on observables change depending on the
input cosmological model.

It is unlikely that straightforward brute force approaches will
succeed in solving this problem. A number of open areas offer
promising directions where progress can be made. These include optimal
data compression, new estimation techniques, use of emulators, and
simplified simulation methods. Although it is too early to pedict the
actual number of large-scale simulations needed, it would appear that
in the exascale era, assuming advances in the actual statistical
techniques mentioned, the simulation resources will be sufficient for
this task (roughly three orders of magnitude more than currently
available) .

\subsubsection{Supernova Simulations}

Observations of Type Ia SNe have led to an empirical relation between
the peak brightness of these events and their decay time, providing a
way to accurately calibrate their intrinsic brightness and, therefore,
their distance. Future precision cosmology missions (e.g. WFIRST) will
require calibration of the brightness-width relation to a much tighter
tolerance than is currently possible with tuned, parameterized
explosion models. Large-scale computer simulations of Type Ia
supernovae therefore have an important role to play in enabling
astronomers to use these explosions to determine the properties of
dark energy.  Such simulations have led to a better understanding of
these events, revealing systematic effects that must be considered
when using them as ``standard candles.'' Only by performing reliable
explosion simulations can systematic biases due to factors such as
progenitor mass and chemical composition, and viewing angle, be
extensively studied, and hopefully removed. However, turbulent
thermonuclear combustion is inherently multidimensional, requiring
explosion simulations in three spatial dimensions. This, coupled with
the need to evolve nuclear reaction networks of requisite complexity
at each grid point, means that Type Ia SNe simulation stresses the
world's largest computational platforms and will continue to do so all
the way to the exascale and beyond.

Experiments and observations have revealed many of the properties of
neutrinos -- mass-squared differences, and three of the four
parameters in the vacuum unitary transformation between the neutrino
mass (energy) states and the weak interaction eigenstates (flavor
states), and only the CP-violating phase remains to be
measured. However, the absolute neutrino masses, as well as the
neutrino mass hierarchy, remain unknown. Aside from cosmological
probes, core collapse supernovae and compact object merger events are
uniquely sensitive to neutrino flavor mixing and neutrino mass
physics. Recent large-scale numerical simulations have revealed that
the neutrino flavor field in these environments can experience
collective neutrino flavor oscillations which can affect the expected
neutrino burst signal from a detected core collapse event in the
Galaxy and can affect issues in energy transport and
nucleosynthesis. If detected, the fossil features of collective
oscillations, i.e., could tell us the neutrino mass hierarchy; and,
conversely, measurement of the hierarchy in the lab makes the
supernova and compact object simulations more reliable and
predictive. However, the approximations underlying the current
paradigm for supernova neutrino modeling are suspect. The first forays
into complete quantum kinetic approaches to following neutrino flavor
evolution suggest that there may be surprises. The stakes are high, as
the compact object and core collapse environments are the key
cosmological engines of element synthesis.

\subsubsection{ Impact of Technology Developments}

Changes in the hardware and software environment~\cite{microproc}
associated with petascale computers are complicating the overall
simulation effort for both N-body and hydrodynamic simulations. In
particular, baryonic effects are associated with extreme spatial and
temporal dynamic range requirements and an increase in the number of
solution variables. Typically the former is addressed using adaptive
mesh refinement or Lagrangian particle techniques which limit high
spatial resolution to locations that require it. The latter is a more
fixed requirement: while algorithms vary in terms of the number of
auxiliary variables they require, they must follow at least density,
momentum, energy, and composition variables. Magnetic fields and
relativistic species introduce additional required
variables. Radiation transport is especially demanding in this
respect. Therefore, cosmological simulations including baryonic
effects are typically memory-limited. This is problematic given
industry projections that coming generations of computers will be
severely unbalanced, with exaflop machines likely to have at most tens
of petabytes of memory. Exascale cosmological simulations including
baryonic effects will need to rely heavily on adaptive resolution
techniques in order to accommodate the required number of solution
variables. A dedicated effort to improve the parallel scalability of
these techniques is therefore required.

The reliance of newer high-performance computers on multiple levels of
parallelism also poses a challenge for baryonic simulations -- and an
opportunity. Typically additional levels of parallelism are
accompanied by high communication latencies between levels,
encouraging separation of tasks between CPUs and accelerators such as
GPUs. This greatly complicates parallel code development; widespread
exploitation of these architectures is currently hindered by the lack
of a common programming model (such as MPI provided for parallel
computers). However, strides are being made toward such a model, and
if we assume that one emerges, it becomes possible to see how multiple
levels of parallelism can be turned to an advantage: by matching the
structure of the problem to the structure of the computer. As noted
above, improved subresolution models are an important component of
current development in cosmological hydrodynamic simulations. To be
mathematically well-posed, these models rely on a separation of length
scales. Below a characteristic scale, the physics incorporated into
the model must be independent of long-range interactions, depending
only on local conditions. Above this scale, long-range interactions
and boundary effects require the use of direct simulation. Therefore,
computers with multiple levels of parallelism may best be exploited by
employing the more traditional levels (networked CPUs) for direct
simulation of the hydrodynamic effects associated with large scales,
while exploiting accelerators for subresolution effects that do not
require long-range coupling.

\section{Data-Intensive Science Challenges}

The challenge posed by the large data stream from experiments requires
a number of responses in scalable data analytics, data-intensive
computing, networking, and new methods to deal with new classes of
research problems associated with the size and richness of the
datasets. The Cosmic Frontier experiments have their commonalities and
differences with traditional HEP experiments, thus the associated data
management strategy has to be independently developed, while maximally
leveraging capabilities that already exist.

\subsection{Data Growth in Experiments and Simulations}

There is a continued growth in data from Cosmic Frontier
experiments. Survey experiments currently exceed 1PB of stored
data. Over the next decade the mass of data will exceed 100PB. In
subsequent decades the development of radio experiments and energy
resolving detectors will result in an increase in data streaming to $>
15$GB/s. Simulation requirements are also projected to increase
steeply. Current allocations are estimated to be of the order of 200M
compute hours/year, with associated storage in the few PB range, and a
shared data volume of the order of 100TB. Data management standards
and software infrastructure vary widely across research teams. The
projected requirements for 2020 are an order of magnitude improvement
in data rates (to 10-100GB/s), a similar increase in peak
supercomputer performance ($~200$PFlops), and the ability to store and
analyze datasets in the 100PB class. It is difficult to make precise
estimates for 2030 as hardware projections are hazy, however, the
science requirements based on having complete datasets from missions
such as LSST, Euclid, and large radio surveys would argue for at least
another order of magnitude increase across the board.

\subsection{Data Preservation and Archiving}

Archiving the observational data in of itself does not appear to be a
huge challenge, as long as it is understood that each dataset needs
to be stored redundantly, preferably in a geoplexed way. The projected
data volumes involved are not particularly large compared to
commercial datasets (with the possible exception of SKA). Given that
the eventual data volumes will probably exceed a few exabytes, the
analyses must be co-located with the data.

The most likely high-level architecture for scientific analyses will
be a hierarchy of tiers, in some ways analogous to the LHC computing
model, where the Tier 0 data is a complete capture of all raw data,
but then derived and value added data products are moved and analyzed
further at lower tiers of the hierarchy, which are not necessarily
co-located with the Tier 0 data centers.

The archives will have to be based upon intelligent services, where
heavy indexing can be used to locate and filter subsets of the
data. There is a huge growth in the diversity of such ``Big Data
Analytics'' frameworks, ranging from petascale databases (SciDB,
Dremel, etc.) to an array of NoSQL solutions. Over the next 5 years a
few clear winners will emerge, allowing the research community to
leverage the best solutions. A high speed, reliable and
inexpensive networking infrastructure connecting the instruments and
all the sites involved in the archiving will be crucial to the success
of the entire enterprise.

\subsection{Computational Resources for Experiments}

The use of computational resources will need to grow to match the
associated data rates for the processing and analysis of observational
data and for simulated astrophysical and cosmological processes. Most
(but not all) of the data processing pipelines use linear time
algorithms, thus the amount of processing is roughly proportional to
the amount of data collected by the instruments. Exceptions to the
linear law will be algorithms which will incrementally reprocess all
the data from a given instrument over and over, whose processing
capabilities must therefore grow as a quadratic function of time.

Most pipelines can be characterized by the number of cycles needed to
process a byte of data. Typical numbers in astrophysics today range
from a few thousand to 100K cycles, thus to process a canonical 100PB
dataset, 10$^{22}$ cycles, or about a billion CPU hours, are
required. One particular characteristic of this processing is that it
will require a reasonable, but not excessive sequential I/O rate to
the disks the data is stored on, typically less than a GB/s per
processing node.

Much of this processing is massively parallel, and thus will execute
very well on SIMD architectures. Emerging many-core platforms will
likely have a huge impact on the efficiency of pipeline
computing. While these platforms are harder to code against, pipeline
codes will be based on well-architected core libraries, where it will
be cost efficient to spend resources to optimize their parallel
execution, thus substantially decreasing the hardware investment.

\subsection{Computational Resources for Simulations}

The required computational resources for simulations naturally fall
into three tiers. The first level of analysis takes place in the host
supercomputer, the second on data stored in the file system. In the
near future, the second step is likely to evolve to an `active
storage' model where large-scale dedicated computing resources may be
accessible, either embedded within storage, or accessible by fast
networking. Finally, the last level of analysis will take place on
object catalogs, generated by the previous two levels. While the first
two levels can be accessed in batch mode, the last level should allow
for interactive access. Finally, since simulation and observational
data are likely to be very similar as simulation fidelity continues to
improve, the last level should also support the same pipelines as
those employed by the experiments.

The volume of data generated by simulations is limited only by
available storage. The actual data volume, however, is controlled by
science requirements, and does not have to hit storage limits. Storage
estimates range from $\sim$1PB in 2013 rising to $\sim$20PB in 2018,
with an associated computation requirement rising from thousands of
cores to $\sim 10^5$ cores (with a core considered to be equivalent of
a current generation microprocessor core).

\subsection{New Computational Models for Distributed
    Computing}

Today's architectures for data analysis and simulations include
supercomputers, suitable for massive parallel computations where the
number of cycles per byte of data is huge, possessing a large
distributed memory, but with a relatively small amount of on-line
storage. Database servers occupy the opposite range of the spectrum,
with a very large amount of fast storage, but not much processing
power on top of the data. For most scientific analyses the required
architecture lies somewhere in between these two: it must have a large
sequential I/O speed to petabytes of data, and also perform very
intense parallel computations. 

Fast graph processing will become increasingly important as both large
and complex simulations are analyzed and as one tracks complex
spatio-temporal connections among objects detected in multi-band
time-domain surveys. To efficiently execute algorithms that require
large matrices and graphs, it is likely that a combination of large
(multiple TB) memory (RAM) will be combined with multiprocessors to
minimize communication overhead. Also, new storage technologies with
fast random access (SSD, memory bus flash, phase change memory, NVRAM)
will play a crucial role in the storage hierarchy.

\subsection{Data Analytics Infrastructure for
    Experiments and Simulations}

One may envision the development of the necessary infrastructure as
building a novel microscope and/or a telescope for data. We need a new
instrument that can look at data from a far perspective, and show the
``big picture'' while allowing the smallest details to be probed as
well. Considering the challenges from this perspective leads to the
understanding that there are similar engineering challenges in
building the shared analytics infrastructure as there are in building
a new experimental facility.

Large-scale datasets, arising from both simulations and experiments,
present different analysis tasks requiring a variety of data access
patterns. These can be subdivided into three broad categories.

Some of the individual data accesses will be very small and localized,
such as accessing the properties of individual halos, or galaxies, and
recomputing their observational properties. These accesses typically
return data in small blocks, require a fast random access, a high IOPS
rate and are greatly aided by good indexing. At the same time there
will be substantial computation needed on top of the small data
objects. These accesses can therefore benefit from a good underlying
database system with enhanced computational capabilities. Going beyond
the hardware requirements, this is an area where the clever use of
data structures will have an enormous impact on the system
performance, and related algorithmic techniques will be explored
extensively. The challenge here is that the small data accesses will
be executed billions of times, suggesting a parallel, sharded database
cluster with a random access capability of tens of millions of IOPS
and a sequential data speed of several hundred GB/s, with an unusually
high computing capability inside the servers themselves.

At the other end of the spectrum are the analyses that will have to
touch a large fraction, possibly all of the data, like computing an
FFT of a scalar field over the entire volume, or computing correlation
functions of various orders, over different subclasses of
objects. These require very fast streaming access to data, algorithms
that can compute the necessary statistics over (possibly multiple)
streams, and hardware that can handle these highly parallelizable
stream computations efficiently (multiprocessors). Here the
requirements would be a streaming data rate in access of 500GB/s
between the data store and the processing, and a peak processing
capability of several PFlops over vectorizable code. These patterns
map best onto traditional HPC systems, with the caveat of the extreme
data streaming requirements.

The third type of access pattern is related to rendering computer
graphics. These tasks will generate various maps and projections,
touching a lot of data, and typically generating 2D images. Such tasks
include computing maps of dark matter annihilation in large
simulations with trillions of particles, ray-tracing to compute
gravitational lensing over a large simulation, ray-traced simulated
images for future telescopes, based on simulations and detailed
telescope and atmospheric models. As many of these tasks are closely
related to computer graphics, mapping to GPU hardware will be very
important, as this approach can yield performance gains of well over
an order of magnitude.

The accuracy requirements of these
computations will also demand that the simulations save larger than
usual numbers of snapshots, increasing the storage requirements. Large
memory machines combined with massive local GPUs are likely to be an
optimal platform for these computations: data can be prefetched in
large chunks into memory and local SSD, and rendered using the
multiprocessors over the local backplanes. Multiple machines can take
different parts of the output and run in parallel.

Dealing with each of these access patterns demands substantial
investments in hardware and software development. To build an
efficient streaming engine, every one of the bottlenecks, both in
hardware and software, must be eliminated as a single chokepoint can
seriously degrade the performance of the whole system. In terms of
algorithms, many traditional RAM-resident algorithms must be recast
into streaming versions. A rethink of statistical algorithm design is
needed, and computations (and computability) should be explicitly
included into the cost tradeoffs.

The need for better programming models, and better high-level
abstractions is evident. In a complex, massively parallel system it
will become increasingly difficult to write code explicitly
instructing the hardware. Therefore, there is a need to explore and
embrace new declarative programming models where the explicit
execution of the code is transparent to the use. At a higher level,
there is a pressing need for the development of a sustainable software
effort that can provide a baseline of support to multiple experiments,
with experiment-specific extensions being built on top of such a
capability. This will require a community effort in the development
and exploitation of new algorithms, programming models, workflow
tools, standards for verification, validation, and code testing, and
long-term support for maintenance and further development of the
resulting software base.

In this new era of computational and data-driven science it is clear
that software is increasingly becoming the major, capital investment,
and hardware (both computational and experimental) is becoming the
disposable, replaceable part, largely due to rapid changes in
silicon-based technologies (bigger CCDs, flash, NVRAM, energy
sensitive photon detectors).  This in turn means, that investments in
software will have a much larger impact than ever before.

\section{Community Issues}
\subsection{Creation of Career Paths}

Over the coming decade Cosmic Frontier science will become ever more
dependent on developments in computing; from the management of data
through to the development of the algorithms that will produce
fundamental breakthroughs in our understanding of the ``Dark
Universe''. This dependence will impose substantial requirements on
the quality and sustainability of the software required by Cosmic
Frontier experiments. There is, however, a growing disconnect between
the way physicists are trained for the research problems of the next
decade and the skill sets that they will require to be
successful. Part of this problem lies in the traditional physics
curriculum and part in the lack of career paths (including tenure
stream careers) of researchers who work at the interface of computing
and physics. Physicists with the skills and activities that satisfy
the computational needs of Cosmic Frontier experiments do not map well
to the traditional success metrics used in academia. Addressing these
needs by the Cosmic Frontier community will require the development of
academic and career mentors for computational physicists, and the
opening of long-term career paths both at national laboratories and at
universities. A number of areas of HEP (e.g. the Energy Frontier) have
addressed many of these challenges and may provide exemplars for the
Cosmic Frontier community.

\subsection{Community Approach to Scalable and Robust
    Software}

The robustness of methods and software implementations is a major
issue that must be considered by the entire community. Best practices
from software enginering should be imported into the next generation
of simulation and analysis codes and workflows, and in modernizing
some of the current workhorse tools. Additionally, there is
a need to have multiple codes and frameworks that are continuously
cross-validated against each other, especially in the context of
error-sensitive analyses, to help isolate errors and bugs of various
kinds. Several public examples of code and analysis tests and
comparisons already exist. In all cases, lessons have been learnt,
some expected, and some unexpected. At the other extreme, there is no
reason to have independent efforts on a very well-defined and specific
task that can be carried out by a smaller group.

This requires the community to come together and define tasks such as
blind challenges and comparisons of critical codes as well as the
timelines for these tasks. Because a significant effort is required in
this area, sufficient resources will have to be committed to this
activity, such as a long-lived testing facility, which may include a
public software registry.

\section{Conclusions}

Lying at the interface between particle physics, cosmology, and
astrophysics, the Cosmic Frontier addresses many fundamental questions
in physics today; from dark energy, to quantum gravity, to the
astrophysics of galaxy formation.  With a 1000-fold increase in data
rates expected over the next decade from a diverse range of Cosmic
Frontier experiments, computational resources and facilities will need
to grow to support the generation, processing and analysis of data
from these experiments. In this era of data intensive science,
simulations, of similar or large scales, will be required to support
the optimization of experimental surveys as well as the physical
interpretation of the experimental results.

Many of the techniques utilized by Cosmic Frontier collaborations are
common across different experiments and simulation frameworks. Most
experiments have, however, developed their analysis and processing
software independently of other programs. There is a growing need for
the development of sustainable software that can provide a baseline of
computational and analytic support for multiple experiments and
collaborations.  This will include both the development of
computational architectures and tools, as well as the software for
analyzing and interpreting the data.  To accomplish this will require
a coherent community effort for long-term support to develop and
implement new algorithms, programming models, workflow tools, as well
as standards for verification, validation, and code testing.

Coupling the computational and data requirements necessitates a
science community who can work at the interface of science, computing,
and data; software will become the instrumentation of the next
decade. Sustained support for the education of the community in the
development and application of these facilities (including the
creation of long term career paths) is of importance to the health of
computing in Cosmic Frontiers.

\end{document}